\documentclass{elsart3}
\usepackage{epsfig}

\usepackage{amssymb}

\newcommand{\lhc}{{\sc Lhc~}}
\newcommand{\slhc}{{\sc SLhc~}}
\newcommand{\vlhc}{{\sc VLhc~}}

\begin{document}

\begin{frontmatter}


\title{Particle Identification\\ for Physics beyond the LHC}

\author{Marco Battaglia}

\ead{marco.battaglia@cern.ch}

\address{CERN, Geneva, Switzerland and University of California, Berkeley, USA}

\begin{abstract}
Accelerator physics beyond the \lhc is expected to provide precision in the study of 
new physics processes which the \lhc may have already unveiled and to extend the 
high energy frontier beyond its reach, in the multi-TeV domain. In this paper 
I review the anticipated needs in terms of particle identification of this physics 
program in relation to the experimental conditions. 
\end{abstract}

\end{frontmatter}

\section{Introduction}
\label{sec:intro}

Charged particle identification has been, and still is, crucial for detecting 
new signals and performing precision measurements at collider experiments. 
At LEP and SLC the identification of leptons has been essential. That of hadrons 
has proven to be beneficial to carry out the physics program in full.

The focus on heavy flavour physics and the study of CP violation in the $B$ sector 
further underlines the need for kaon and proton tagging. $B$ experiments are all equipped 
with Cherenkov detectors to tackle problems such as the discrimination of penguin-mediated 
$b \to s$ transitions from tree-level $b \to u$ processes. At the \lhc proton collider, 
the hadron identification capabilities are 
similarly emphasised by the specialised experiments. Lepton identification is an 
essential component in the design of multi-purpose experiments, from the Level-1 
trigger stage. 

The \lhc is expected to provide a major breakthrough in particle physics, by probing the 
Higgs sector and testing new physics beyond the Standard Model (SM), possibly manifesting the 
existence of Supersymmetry. But to understand the mechanism of origin of mass and electro-weak 
symmetry breaking, neither the study of the Higgs profile nor the search for additional Higgs 
particles will be completed at the \lhc. Some measurements are limited in accuracy while others 
may not be feasible at all. In particular, understanding Supersymmetry requires to identify and 
measure all the supersymmetric particle partners. More generally, given a set of signals at 
\lhc, their nature and properties may not be unambiguously determined, unless additional data is 
provided by a new generation of particle colliders. This paper addresses some of the issues in 
particle identification which emerge from the anticipated physics programs of future colliders, 
with emphasis on $e^+e^-$ linear colliders.

\section{Accelerator Designs and Opportunities}
\label{sec:accel}

There are several alternative paths for reaching the high energy frontier, beyond the 
\lhc. The luminosity in proton collisions can be increased. The \lhc can be upgraded 
to provide collisions at 10$^{35}$~cm$^{-2}$s$^{-1}$. Such upgrade (\slhc) could extend the mass 
reach to new particles by about 30\%, provided the ATLAS and CMS detectors can be re-fitted to 
cope with the higher particle fluxes. But past $10^{35}$~cm$^{-2}$s$^{-1}$, the \lhc detector 
performance greatly degrades, even with major upgrades~\cite{Gianotti:2002xx}. 
At these luminosities, particle 
identification techniques such as transition radiation detectors appear to be not applicable. 
An increase in $\sqrt{s}$ should be significantly easier to be exploited by the experiments. 
This makes a very large hadron collider (\vlhc) a possible long-term option for accelerator 
HEP~\cite{Ambrosio:2001ej}. 
A large $\sqrt{s}$ increase from the \lhc energy has to be supported by very significant R\&D for 
magnets and vacuum. An $e^+e^-$ linear collider (LC), providing collisions at energies from the $Z^0$
 pole up to approximately 1~TeV with luminosity in excess to 10$^{34}$~cm$^{-2}$s$^{-1}$, is 
presently considered as the best motivated option for the next large scale project in accelerator 
particle physics. The linear collider R\&D is reaching the time at which a decision on the technical 
feasibility of the project and an informed choice of the most advantageous technology can be 
taken~\cite{Loew:se}.
Such collider would not directly test the energy range beyond the \lhc, but it will complement its 
data with the accuracy characteristic of lepton colliders. Through the collision of point-like 
particles with tunable energies, the democratic production of particles, including those interacting 
only weakly, and the possibility to modify their polarisation, as well as to replace $e^+ e^-$ with 
$\gamma \gamma$ and $e^- e^-$ collisions, the high energy linear collider offers a unique 
opportunity. Lepton collisions are not limited to the TeV frontier. Active R\&D programs are already 
addressing the challenges offered by multi-TeV collisions with electrons (CLIC) and, in a longer 
perspective, possibly also with muons, and providing conceptual designs.
Hadron and Lepton Colliders give an healthy complementarity of approaches and technological 
challenges which are being addressed by a significant, continued world-wide R\&D effort. 

\section{Physics Program and Particle Identification}
\label{sec:phys}

The details of the physics program for the next generation of particle colliders 
will be defined by the exploration of the TeV frontier by the \lhc. Our present 
understanding indicates that the study of the Higgs sector will require efficient 
flavour tagging, to determine its quark couplings, efficient lepton identification 
for mass, spin and gauge couplings measurements. Several models of new physics which 
address the hierarchy problem are characterised by a rich spectroscopy of new 
states, with abundance of multi-lepton signatures.

The \slhc will requires major tracker rebuilds to maintain jet tagging capabilities and 
fully benefit from the higher luminosity. Moving to higher energies poses problems which 
become particularly severe in the forward regions where jet and lepton tagging are essential. 

\begin{center}
\begin{table}
\caption{Track density and tracker occupancy (normalised to \lhc) in proton collisions 
from the \lhc to the \vlhc.}
\begin{tabular}{|l|c c c c|}
\hline
            & \lhc & \slhc & \vlhc-I & \vlhc-II \\
\hline
$\sqrt{s}$ (TeV) & 14   &  14   &  40     &  200 \\
$dN/d\eta$/BX & 150 & 750 & 180 & 500 \\
$<E_T>$ (GeV)      & 0.5   & 0.5 & 0.5 & 0.6 \\
Tracker Occupancy & 1 & 10 & 1 & 3 \\
\hline
\end{tabular}
\end{table}
\end{center} 

In this environment lepton identification needs to be guaranteed as multi-lepton 
signatures will represent an important handle in isolating signals of new physics. 
A good example is offered by $WW$ and $ZZ$ vector boson scattering. If no elementary 
Higgs boson exists, electroweak symmetry breaking should happen dynamically. The effects 
from strong boson scattering should become observable at the \slhc with good statistical 
significance analysing the leptonic decays of the gauge bosons. These studies will have 
to rely on efficient and well modelled response functions for lepton identification in 
the high luminosity environment.

The $e^+e^-$ linear collider offers more benign experimental conditions to 
consider advanced particle identification techniques with a wide array of physics 
opportunities and aims at high precision measurements. It thus offer 
important options for characterising the particle identification needs 
of collider physics beyond the LHC.
Electron identification can be obtained from the response of the finely segmented 
electromagnetic calorimeter. Muons are identified in the instrumented return flux iron. 
Plastic streamer tubes, or RPC detectors, can be used to provide multiple samplings since rates 
are low. The typical momentum of leptons in muons in jets is $\simeq$~20~GeV, corresponding 
to a lateral spread of $\simeq$~2~cm after 7~$\lambda$, which sets the scale of the needed 
position resolution. Assuming a solenoidal magnetic field of 4~Tesla, the muon momentum cut-off 
is $\simeq$~5~GeV for a coil located at 4~m from the interaction region.

\begin{figure}
\begin{center}
\epsfig{file=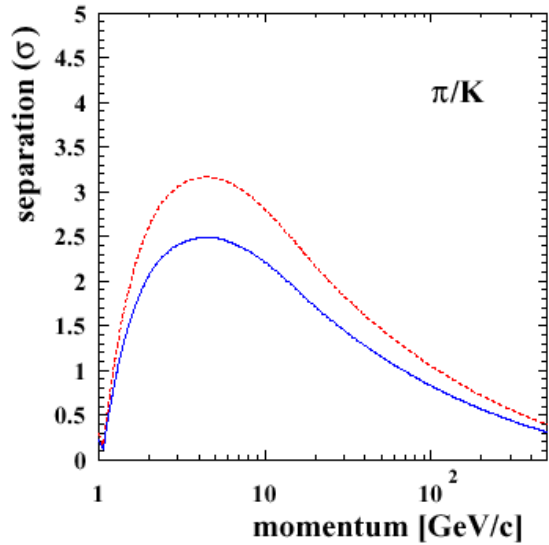,width=7.5cm}
\end{center}
\caption{Hadron identification with $dE/dx$ in TPC at the Linear Collider: $\pi$/$K$ separation 
as function of particle momentum in units of $\sigma$ for the proposed TESLA detector 
(from~\cite{Behnke:2001qq}).}
\label{fig:dedx}
\end{figure}

Charged particle identification offers crucial sensitivity to new phenomena 
and provides redundancy in discriminating signals for precision measurements.
In principle, it would be advantageous to provide the LC experiment with both lepton 
and hadron tagging capabilities. However, analysing the multi-jet final states requires 
excellent parton energy and direction reconstruction. This can be best achieved with 
the energy flow technique, which combines tracking and calorimetric information. There 
is a need to minimise the material in front of the electro-magnetic calorimeter which imposes 
some compromises. In absence of a clear case for hadron identification with good purity,  
current detector designs are relying on the information which can be obtained from specific 
ionization, dE/dx, in the Main Tracker device for hadron identification. Both a large 
volume Time Projection Chamber (TPC) and a Si tracker are being considered. The TPC design 
included in the TESLA proposal has 240 points/track with micro-pattern gaseous detector read-out. 
A dE/dx resolution of $\simeq$~4~\% is achievable, corresponding to a $\ge$~2.0~$\sigma$ 
$\pi$/$K$ separation in the kinematic region of 1.5~GeV $< p <$ 20~GeV 
(see Figure~\ref{fig:dedx})~\cite{Behnke:2001qq}. 
Together with the excellent mass resolution afforded by the accurate momentum determination, 
this would ensure a good separation of processes such as $B \to \pi \pi$ at a Giga-$Z$ factory.
A discrete Si Tracker may provide dE/dx measurements in Si, as pioneered by DELPHI and Belle. 
However, the resolution is much poorer, due to limited number of samplings available, and 
hadron tagging would only be possible in the $1/\beta^2$ region, since the relativistic rise 
saturates.

Further specifications for hadron identification may come from physics at the $W^+W^-$ 
threshold where some of the elements of the CKM mixing matrix could be directly determined, 
with interesting accuracy if $c$ and $s$ jets can be efficiently tagged. Such direct 
measurements will be free from the theoretical uncertainties and model dependence 
which affect the determination derived from partial decay widths. This would call for 
good $K$ identification over a wide momentum range. However, at higher energies benefits 
from $c$ and $s$ tagging in $W^{\pm}$ boson decays, appear to be marginal. The study of 
$W_LW_L$ production represents a good test case. An increase of longitudinally polarised 
$W$ pair production in high energy $e^+e^-$ collisions is predicted some models, including 
technicolor. The helicity of $W$ bosons can be best measured in mixed 
$e^+e^- \to WW \to \ell \nu c \bar s$ decays where the lepton charge and the flavour of 
the $c$ quark can be used as analysers. Charged kaon identification enhances the tagging 
quality. But including realistic performances dilutes such improvement to only 
$\simeq$~20~\%~\cite{Soffer:2001ce}. 
However, we should remain aware that the most exciting part of the physics program at future 
colliders may come from physics which we cannot anticipate today. Therefore the detector 
concepts should ensure that some redundancy in the tagging capabilities is kept.
 
The case for excellent lepton identification is easily made, since it is instrumental to some of the 
most fundamental processes on the linear collider agenda. The study of the Higgs boson through the 
signature process $e^+e^- \to Z^0H^0$ relies on tags of the leptonic $Z^0$ decay to perform studies 
independent on the Higgs decay properties. Since these leptons are quite energetic, their 
identification is ensured through the electro-magnetic, hadronic calorimeter and muon chamber 
response. 
But other cases are less straightforward. A detailed study of Higgs boson couplings to fermions and 
gauge bosons, requires the use of inclusive four jet events. Here semi-leptonic $b$-quark decays 
giving will distort the $M_{JJ}$ invariant mass distribution. It is thus important to tag soft 
leptons in jets to apply appropriate corrections. If new physics exists, to cancel the effect of 
heavy Higgs mass in electroweak observables, its mass $M_H$ can be significantly heavier than the 
$\simeq$~210~GeV limit derived from present data. In this case it would become interesting to search 
for $e^+e^- \to H^0 e^+e^- \to X e^+e^-$ at high energies, which has a large cross section and 
allows model-independent analyses to be performed by tagging the forward electrons. The $ZZ$ fusion 
analysis needs to identify electrons and precisely measure their energy and direction, down to 
$\simeq$~100~mrad, close to the bulk of the $\gamma \gamma$ background.

If Supersymmetry is realised in nature, the $e^+e^-$ collider is also expected to complement the 
\lhc to determine its fundamental parameters and establish whether the lightest supersymmetric 
particle (LSP) is indeed responsible for dark matter (DM) in the Universe. In particular, the LC 
will produce and measure accurately the $\tilde{\ell}$ sleptons and gauginos, including the LSP, 
properties. The recent 
WMAP satellite determination of DM density constrains the phase space of supersymmetric model 
parameters. In a constrained MSSM these data indicates that the supersymmetric spectrum is likely to 
have small slepton-LSP mass differences, corresponding to rather soft spectra of the lepton produced 
in $\tilde{\ell}$ decays~\cite{Ellis:2003cw,Battaglia:2003ab}. 
\begin{figure}[ht!]
\begin{center}
\epsfig{file=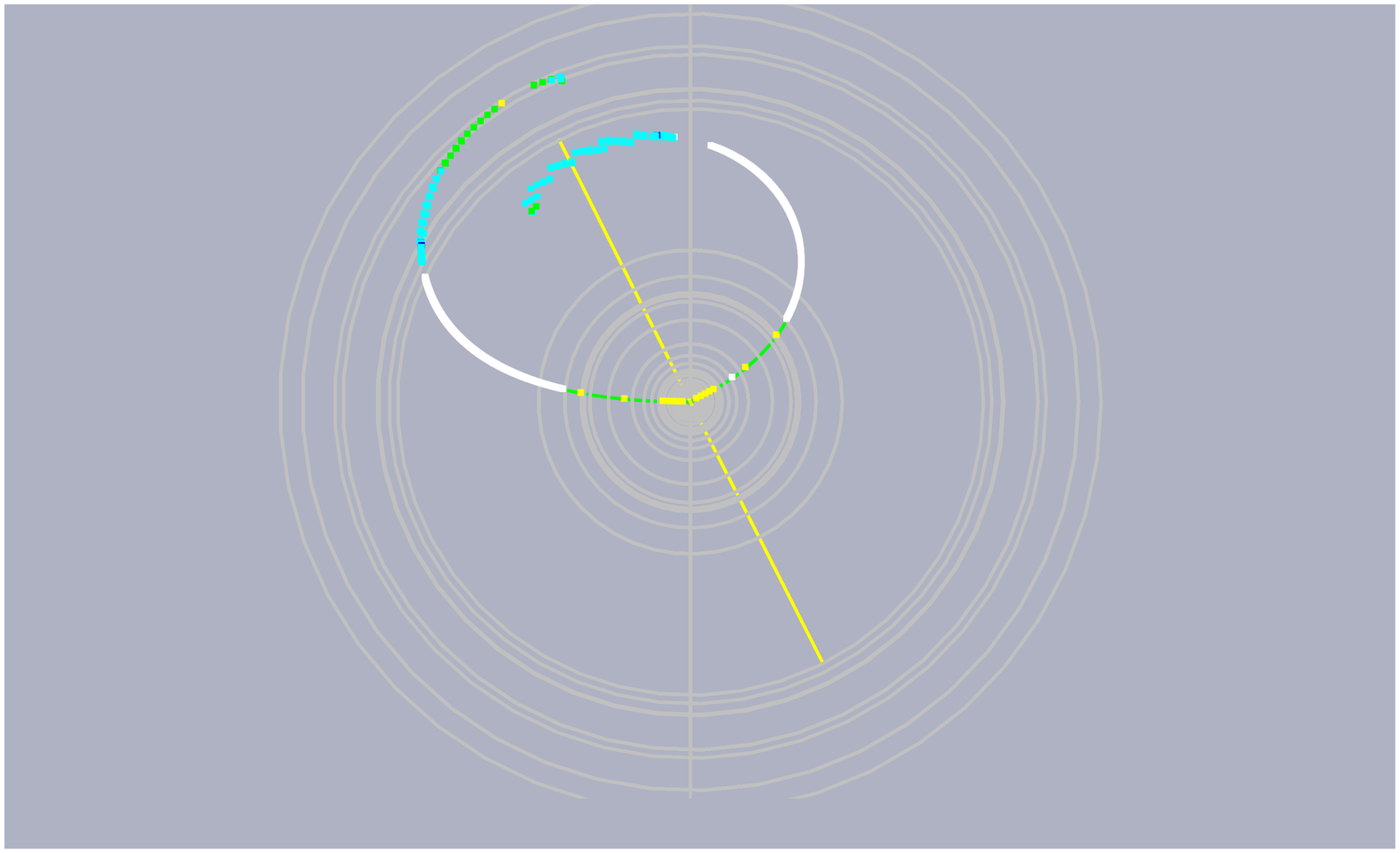,angle=-90,width=7.5cm,clip=}
\end{center}
\caption{Supersymmetric particle production with low momentum leptons: a fish-eye view of a 
$e^+e^- \to \tilde{\mu_R}^+ \tilde{\mu_R}^- \to \mu^+ \chi^0_1 \mu^- \chi^0_1$ event at 
$\sqrt{s}$~=~1~TeV. One muon is produced at the lowest momentum available to the reaction and 
is not reaching the muon detectors.}
\label{fig:evt}
\end{figure}

\begin{figure}[ht!]
\begin{center}
\epsfig{file=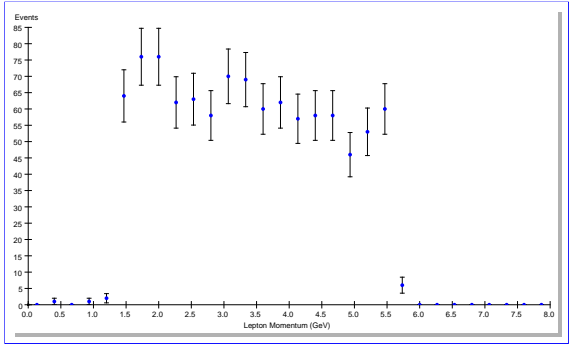,angle=-90,width=7.5cm} 
\end{center}
\caption{Supersymmetric particle production with low momentum leptons: momentum distribution 
of leptons from $e^+e^- \to \tilde{\mu_R}^+ \tilde{\mu_R}^- \to \mu^+ \chi^0_1 \mu^- \chi^0_1$ 
events at $\sqrt{s}$~=~1~TeV. The $\tilde{\ell}$ and $\chi^0_1$ masses can be extracted from the 
location of the lower and upper kinematic endpoints.}
\label{fig:pmuon}
\end{figure}
The slepton and neutralino masses can be accurately determined from the 
endpoints of the lepton momentum distribution in the two-body decay. But at the upper tip of the 
cosmologically interesting region, where the two particles are almost degenerate, the lower lepton 
energy edge may be located at about 1~GeV (see Figures~\ref{fig:evt} and \ref{fig:pmuon}). 
This raises the issue of 
lepton tagging capabilities extending to such low energies matched with efficient event selection 
in presence of accelerator-induced backgrounds.

A similar scenario arises in a particular class of models with extra dimensions. 
Extra dimensions are being actively explored as an alternative solution to the hierarchy problem. 
Most realisations are expected to give spectacular signals at future colliders. Universal extra 
dimension (UED) models have all SM particles propagating in one or more compact extra dimensions. 
Kaluza Klein partners of SM particles have masses $\simeq n R^{-1}$, where $R$ is the 
compactification radius, and identical spin and couplings to SM particles. The mass degeneracy of 
particles of the same KK-level $n$ is only broken by radiative corrections and KK-parity conservation
guarantees that the lightest KK state, generally the photon or neutrino excitation, is stable. 
So UED also offers a viable CDM candidate in the form of its lightest KK 
particle~\cite{Servant:2002aq}. 
This scenario can be tested in details at a LC of sufficient energy through processes 
where lepton KK excitations are pair produced to decay into ordinary leptons and stable lightest KK 
states. This closely resembles the slepton production and decay process in Supersymmetry, except 
that here the mass splitting is always small and measuring the lower lepton energy edge requires 
tagging soft muons and electrons.

Studying $e^+e^- \to f \bar{f}$ at $\sqrt{s} \ge$~1~Tev will provide a window on New Physics at 
scales of order 10~TeV and more, which are much beyond those directly reachable at the next 
generations of colliders. But to ensure sensitivity to new phenomena through the electro-weak fits, 
it will be important to perform accurate forward-backward asymmetry measurements for both leptons 
and quarks, which ensure the best sensitivity at the upper scale edge~\cite{Battaglia:2002sr}. 
The reaction $e^+e^- \to t \bar{t}$ at multi-TeV energies offers an interesting case. Here the charge
of the quark can be determined from that of the lepton produced in the leptonic $W$ decay. Therefore,
indirect sensitivity to the 10~TeV scale will rely on efficient tagging of relatively soft leptons. 

\begin{figure}[ht!]
\begin{center}
\epsfig{file=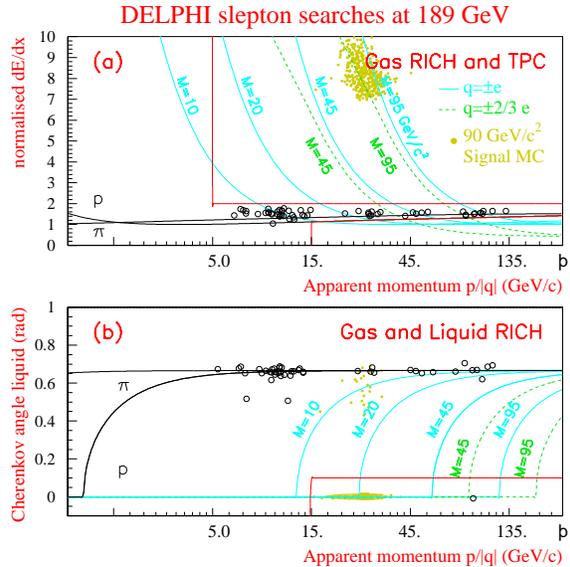,width=7.5cm} 
\end{center}
\caption{Heavy stable particle search with {\sc Delphi} at {\sc Lep} (from~\cite{Abreu:2000tn}).}
\label{fig:heavyp}
\end{figure}
New particles may also appear in the form of heavy stable particles, directly detectable with 
particle identification and time of flight (TOF). 
At a radius of 2~m the time-of-flight of a $\beta=1$ particle is 6.7~ns which can be measured 
with $\simeq 50$~ps accuracy. Since data is integrated over several bunch crossings in a train 
and, depending on RF chosen technique, the bunch spacing $\Delta t_b$ is in the range 1-300~ns, 
TOF must deal with ambiguities. The TPC may be used to resolve the bunch crossing by 
extrapolating to the collision point.
An example is offered in the context of Supersymmetry by Gauge-mediated 
Supersymmetry Breaking Models (GMSB), where the gravitino, $\tilde{G}$ is the LSP. Due to its weak 
coupling, the next lightest particle (NLSP) is long-lived. The $\tilde{\tau}$ NLSP scenario with 
$\tilde{\tau} \to \tilde{G} \tau$ can be investigated in $\tilde{\tau}$ pair production. 
The signature is a pair of heavy long-lived charged particles traversing the detector. 
Their existence can be revealed through TOF and specific ionization, if not dedicated particle 
identification detectors, using a technique pioneered at {\sc Lep} 
(see Figure~\ref{fig:heavyp})~\cite{Abreu:2000tn}. 
Care needs to be taken to ensure that the resulting large ionization signals are preserved by the 
dynamic range of the read-out electronics.

\section{Conclusion}
\label{sec:close}

The future of particle physics depends on the availability of new generations 
of colliders after the \lhc. The $e^+e^-$ linear collider is the next large 
scale project under consideration. While hadron identification does not appear to 
have a major impact on the LC physics capabilities, identification of leptons, and also 
of new heavy stable hadrons, will be instrumental to many of its key physics processes. 
Examples from the main lines of investigations expected at present, indicate the 
challenging requirements from the wide range of kinematics offered by anticipated 
physics signatures.

\end{document}